\documentclass[twocolumn]{aastex62}
\usepackage{amsmath}
\usepackage{amssymb}
\usepackage{graphicx}
\usepackage{multirow}
\usepackage{color}

\graphicspath{{./}{figures/}}

\received{xxx}
\revised{yyy}
\accepted{zzz}
\submitjournal{ApJ}
\shorttitle{Modelling PSD in AGN}
\shortauthors{P. Chainakun}

\begin{document}

\title{Modelling the X-ray power spectra of AGN by using reprocessing echoes from an extended corona}
\correspondingauthor{Poemwai Chainakun}
\email{pchainakun@g.sut.ac.th}
\author[0000-0002-9099-4613]{P. Chainakun}
\affil{School of Physics, Institute of Science, Suranaree University of Technology, Nakhon Ratchasima 30000, Thailand}

\begin{abstract}

Characteristic signatures that X-ray reverberation from an extended corona can manifest in the observed PSD of AGN are investigated. The presence of two X-ray blobs illuminating an accretion disc can cause the interference between two reprocessing-echo components and produce distinct physical features in the PSD. The oscillatory structures (e.g., dips and humps) are seen but, contrarily to the lamp-post case, the strongest dip is not always the one at the lowest frequency. Instead, we find the frequency where the strongest dip is seen associates to the lower-source height while the lowest frequency where the first dip appears links with the upper-source height. This is because the reverberation timescales increase with the source height. Accurate modelling of the PSD then helps put constraints to the lower and upper limit of the corona extent. Furthermore, the reverberation signatures are less pronounced with increasing number of sources that do not produce reflection (e.g., additional X-rays from fast, relativistic outflows). The amplitude of the oscillations also depends on the amount of dilution contributed by the X-ray sources, thus encodes information about their relative brightness. Due to stronger dilutions, robust detection of these signatures with the current observations will become even more difficult if the corona is extended. Future observations made by \emph{Athena} will enable us to fit these characteristics in statistically significant details, and to reveal the nature of the disc-corona system.

\end{abstract}
\keywords{accretion, accretion discs - black hole physics - galaxies: active - X-rays: galaxies}

\section{Introduction} \label{sec:intro}

The X-ray variability is commonly used to study the mechanisms that power accreting black holes and to constrain the innermost region closest to the event horizon. An effective tool to assess fast variability is the power spectral density (PSD) defined as the variability power as a function of the temporal frequency $P(f)$. The typical PSD of the Active Galactic Nuclei (AGN) has a broad band power-law shape with one or two bend frequencies where the profile changes its slope \citep[e.g.][]{Martin2012}. The AGN usually have high variability power at low frequencies, meaning that their largest amplitude variability is on long timescales. The variability power decreases significantly at high frequencies. These properties were also observed in the black hole binaries but with smaller timescales due to a smaller black hole mass \citep[e.g.][]{Cui1997, Remillard2006}, suggesting a similar production mechanism in black hole binaries and AGN. \cite{Martin2012} investigated the X-ray temporal properties of 104 nearby AGN available in \emph{XMM-Newton} archive and found that the majority of variable sources exhibited a PSD that was well fitted by a single power-law with a mean index of $\sim2$ and a mean bend frequency of $\sim2 \times 10^{-4}$ Hz. Propagating of the mass accretion rate fluctuations that modulate the softer, outer regions first before the inner, harder regions can produce observed PSD shape and also the hard lags on a wide range of long timescales \citep[see, e.g.,][]{Arevalo2006}.

Another unique way to probe the innermost region close to the black hole is by measuring reverberation lags associating to the light-crossing time between the direct X-ray photons produced in the corona and back-scattered photons from the disc. All works in attempts at modelling reverberation lags under the lamp-post scenario suggested that the AGN corona is compact and confined within $\sim 10$ gravitational radii ($r_{\rm g}$) from the central black hole \citep{Emmanoulopoulos2014, Cackett2014, Chainakun2015, Epitropakis2016, Chainakun2016}. Reverberation lags can be found in $\sim 50\%$ of variable AGN sources in the \emph{XMM-Newton} archive \citep{Kara2016}. In reality the corona can be extended or outflowing \citep{Wilkins2016, Wilkins2017, Chainakun2017}, rather than being a point source. Moreover, the wavy residuals around the best-fit reverberation lags at high frequencies were detected that could not be fully explained under the lamp-post configuration \citep{Caballero-Garcia2018, Caballero-Garcia2019}. These wavy residuals may be signatures of reverberation under an extended corona framework \citep{Chainakun2017}. 

\cite{Papadakis2016} suggested that there should be signatures of X-ray reverberation, or reprocessing echoes, in the power spectra of AGN as well. However, searching through the PSD profiles of many AGN turned out that these reverberation signatures cannot be easily detected due to a limitation of the signal-to-noise in the data \citep{Emmanoulopoulos2016}. \cite{Chainakun2017} modelled the energy-dependent light curves from two-blob configuration (i.e., assume the corona to be two isolated point sources) and used them to predict the time lags and PSD. The model could reproduce traditional PSD properties such that the harder photons vary more at higher frequencies. A high-frequency dip was found in the PSD profiles and the frequency where the dip is seen is energy independent. The amplitude of the dip is also more prominent in the reflection-dominated band, consistent with the lamp-post cases reported by \cite{Papadakis2016}. Here, we investigate further the effects of X-ray reprocessing echoes on the PSD using the model beyond the standard lamp-post one (e.g., two blobs and outflowing). 

\section{Modelling the PSD} \label{sec:floats}

Let us assume $s_{k}$ is the average count rate over duration $\Delta t$ at time $t_{k}=k\Delta t$,  where $k=1$, 2, 3, ..., $N$. The PSD can be estimated by the modulus squared of the discrete Fourier transform of $s_{k}$ \citep[e.g.][]{Nowak1999,Emmanoulopoulos2013}:

\begin{equation}  
    P(f_{j}) = \left|\sum\limits_{k=1}^N s_{k}e^{{\rm i}2\pi f_{j}t_{k}}\right|^{2} \;,
    \label{eq:psd}
\end{equation}  
where the frequencies $f_{j}=j/(N\Delta t)$ for $j=$1, 2, 3, ..., $N/2$. The minimum frequency is $f_{1}=1/(N\Delta t)$ and the maximum (Nyquist) frequency is $f_{N/2}=1/(2\Delta t)$. We normalize the PSD to the light curve mean squared, as in \cite{Papadakis2016}. All physical quantities throughout the simulations are calculated in natural units ($G=c=1$), and scale with the black hole mass $M$. We set the number of points $N=10^{5}$ and the time bin size is $\Delta t=1 t_{g}$, where $t_{g}=GM/c^{3}\sim4.926(M/10^{6}M_{\odot})$ s.  

\subsection{The model set-up}
\begin{figure}
    \centerline{
        \includegraphics[width=0.45\textwidth]{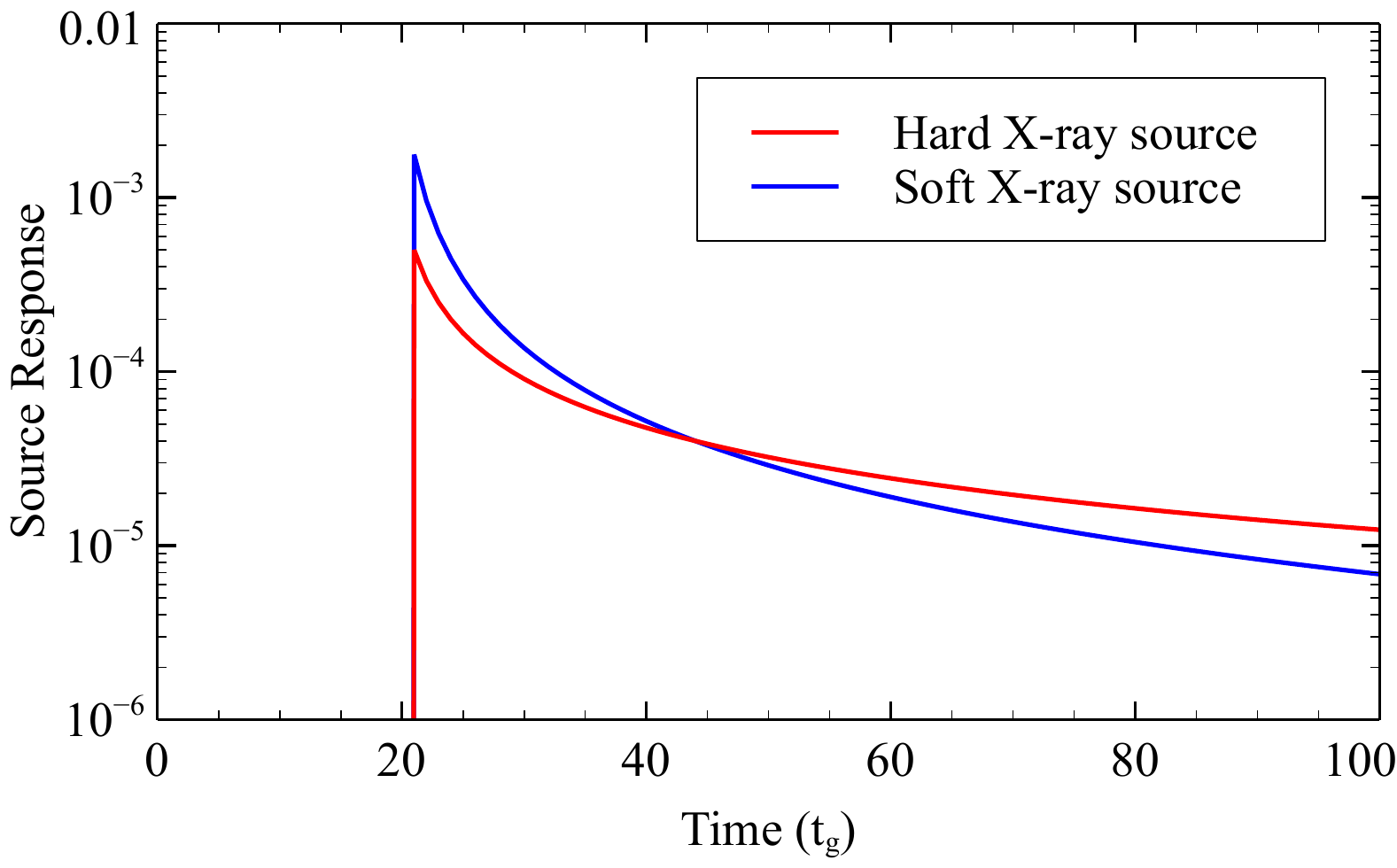}
    }
    \caption{Examples of the source-response functions, $\Psi_{i}(t)$, modelled as cut-off power-laws.}
    \label{p-sres}
\end{figure}

Let us begin with the dual lamp-post case when the accretion disc (assumed geometrically thin and optically thick) is illuminated by two X-ray blobs. The variation of each X-ray source can be triggered by the same primary variation, $x_{0}(t)$, but how each source responds is explained by the source response function, $\Psi_{i}(t)$. The source variability in the energy band $E_{j}$ is the convolution between the primary variation, $x_{0}(t)$, and the source response, $\Psi_{i}(t)$,
\begin{equation}
\begin{split}
 x_{i}(E_{j},t) & = F_{i}(E_{j})\:x_{0}(t) \otimes \Psi_{i}(t)\\
 & = F_{i}(E_{j})  \int_{0}^{t} x_{0}(t^\prime) \Psi(t-t^\prime)dt^\prime. 
    \label{eq:var1}
\end{split}
\end{equation} 
Although there is no preferable choice of the two source-response functions because only their relative difference should matter, these phenomenological functions are needed because the model assumes the variability of the source continuum depends on the source-response function. In this work the source response is modelled in the simplest way using a cut-off power law. Examples are illustrated in Fig.~\ref{p-sres}. The source-response functions are produced in the way that the harder source response dominates the softer source response at late time and on long timescales, the conditions of which can successfully produce both positive hard lag at low frequencies and negative soft-reverberation lags at high frequencies \citep{Chainakun2017}. This is a common profile of time lags in AGN where reverberation is taking place. 

Note that the $\Psi_{i}(t)$ is assumed, for simplicity, to be a function of time only. Therefore, the amplitude of the source variability, $x_{i}(E_{j},t)$, in each energy band is different only in the normalization flux $ F_{i}(E_{j}) \propto 
    \int_{E_{j,\rm low}}^{E_{j,\rm high}} {E_{j}^\prime}^{-\Gamma_i} {\rm d} E_{j}^\prime$, where $\Gamma_{i}$ is the photon index of the source emission spectrum.  

We follow the methods outlined in \cite{Chainakun2017} to compute the full-relativistic disc response function, $\psi_{i}(E_{j},t)$, from ray-tracing simulations. The disc response determines the flux of the reflected flux in the energy band $E_{j}$ that a distant observer will detect, as a function of time. The subscript $i$ is referred to the parameters of the $i^{th}$ source. The source variation produces a delayed accretion disc response, $\psi_{i}(E_{j},t)$. The variability of the reflection flux from the disc is calculated through the convolution term $x_{i}(E_{j},t)  \otimes \psi_{i}(E_{j},t)$. The observed X-ray variability in the energy band $E_{j}$ can be written in the form of
\begin{equation}
\begin{split}
      s_{rev}(E_{j},t) &=  B_{1}\Big{[}x_{1}(E_{j},t) + r_{1}\:x_{1}(E_{j},t)\otimes \psi_{1}(E_{j},t)\Big{]} \\
            & +B_{2}\Big{[} x_{2}(E_{j},t) + r_{2}\:x_{2}(E_{j},t)\otimes \psi_{2}(E_{j},t)\Big{]}\:.
    \label{eq:var2}
\end{split}
\end{equation} 
The first and second squared brackets are the flux contribution from the first, lower and the second, upper X-ray sources, respectively. Each source can contribute both direct and reflection flux into the light curve (i.e., first and second term in each squared bracket). $B_{i}$ is the brightness parameter in the observer's frame and, for simplicity, we set $B_{1}=1$ so that $B_{2}<1$ and $B_{2}>1$ refer to the case when the upper source is seen to be fainter and brighter than the lower source, respectively. Here, we normalize the area under the response functions to 1 and employ $r_1$ and $r_2$ to be the reflection fraction associating to the lower and the upper sources. For a given geometry, the reflection fraction of each source is approximated via the ray-tracing simulations by counting the number of photons that hit the accretion disc comparing to that reach the observer. Therefore, the time-integrated value of the $r_{i} \otimes \psi_{i}(E_{j},t)$ still represents the observed ratio between the reprocessed and continuum photons in the energy band $E_{j}$ due to the $i^{th}$ X-ray source. By setting $r_{1}=r_{2}=0$, the observed X-ray variability in equation~\ref{eq:var2} is reduced to the case when the effects of reverberation are excluded, which is
\begin{equation}
      s_{0}(E_{j},t) = \; B_{1} \:x_{1}(E_{j},t) +B_{2}\: x_{2}(E_{j},t) \;.
    \label{eq:var3}
\end{equation} 
Alternatively, the parameter $r_i$ can be combined to the disc response function because they both depends on the geometry of the disc-corona system.

\subsection{The PSD estimation and results}

To compute the PSD from equation~\ref{eq:psd}, we first produce the simulated light curves in equations~\ref{eq:var2}--\ref{eq:var3} by generating the primary variation, $x_{0}(t)$, which is assumed for initial investigation to be a Gaussian random variate. Of course, we do not know the real function of $x_{0}(t)$, but we assume that each X-ray source and its constituent parts (i.e., continuum and reflection components) respond to this primary variation. The PSD$_{rev}$ is the power estimated through equation~\ref{eq:psd} using the light curve $s_{rev}(E_{j},t)$. The PSD$_0$ is also calculated by using the light curve $s_{0}(E_{j},t)$, which has no reflection flux included. The PSD$_{rev}$/PSD$_0$ then represents the ratio between the PSD that includes and excludes reverberation effects. In \cite{Papadakis2016}, the ratio PSD in the lamp-post case was computed directly by the product of the intrinsic PSD with the disc response function. Although we use the different approach by generating variability from different source components, the ratio plots between the PSD$_{rev}$ and PSD$_0$ here should represent the ratio of the observed over the intrinsic PSD as well. Regardless of the intrinsic PSD shape that we do not exactly know, the features imprinted in the PSD$_{rev}/$PSD$_0$ profiles should be purely due to reverberation since they are the ratio comparing to the reference PSD$_0$.    

\begin{figure}
    \centerline{
        \includegraphics[width=0.45\textwidth]{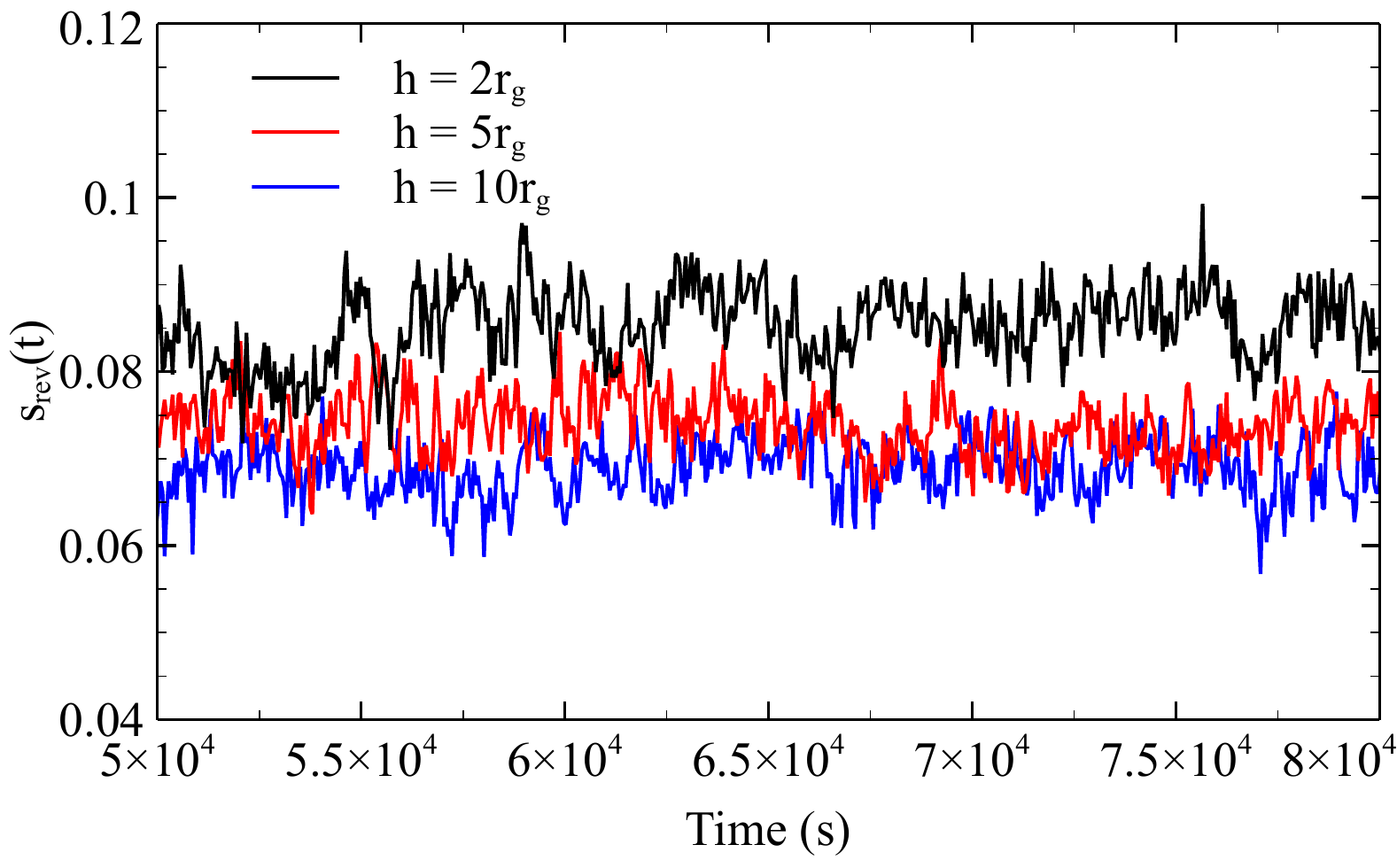}
    }
    \vspace{0.2cm}
    \centerline{
        \includegraphics[width=0.45\textwidth]{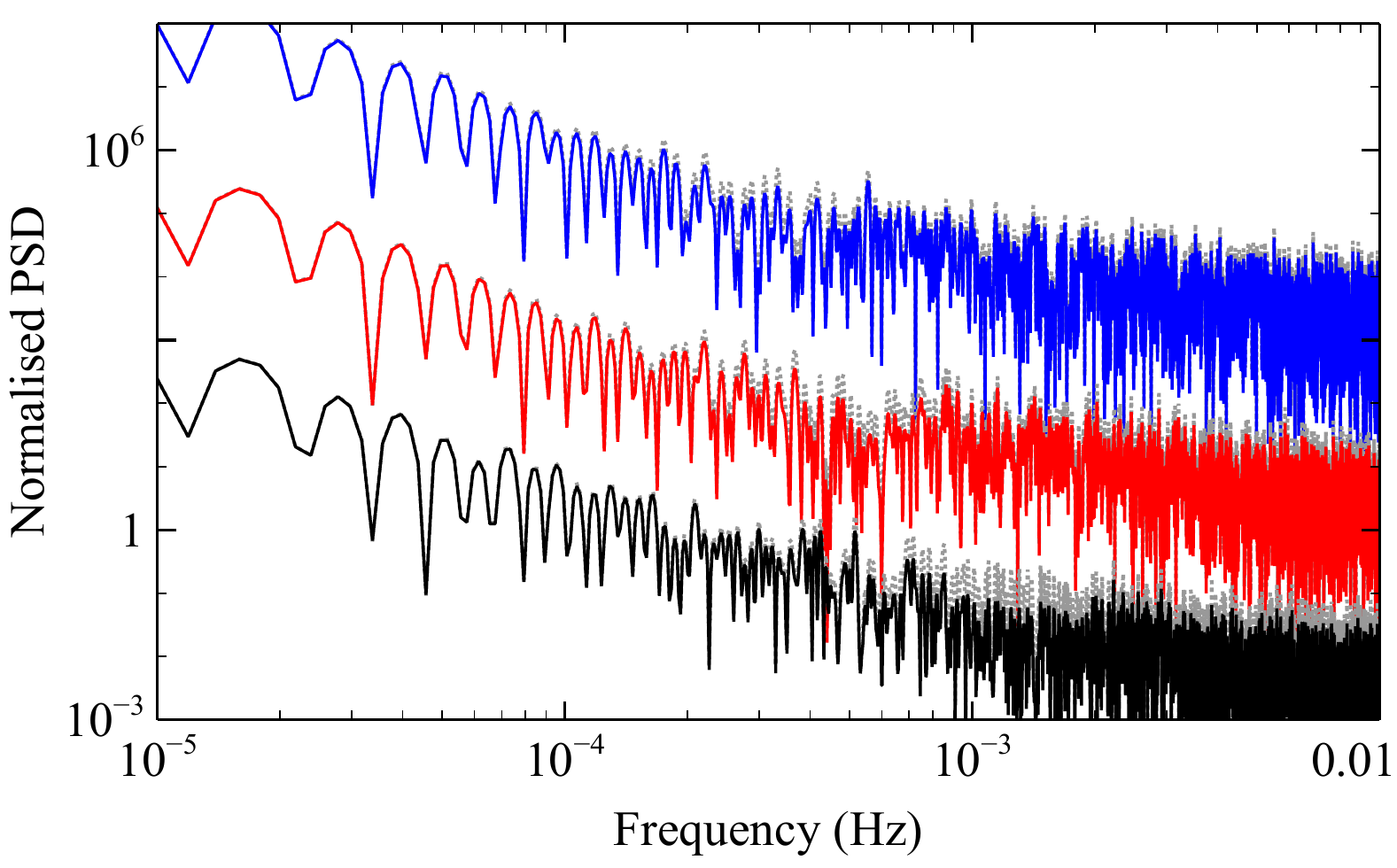}
    }
    \vspace{0.2cm}
     \centerline{
        \includegraphics[width=0.45\textwidth]{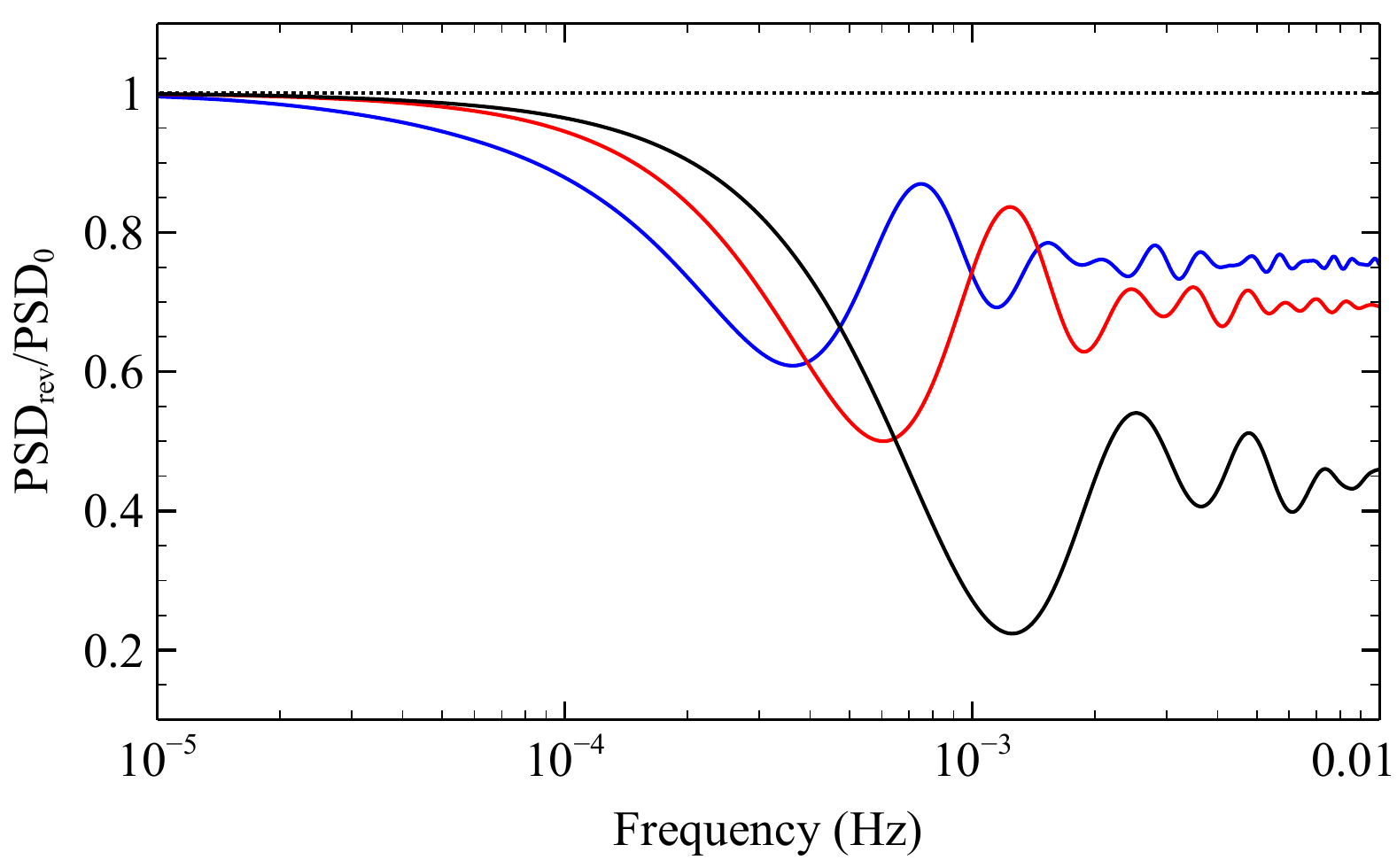}
    }
    \caption{Top panel: simulated light curves $s_{rev}(E_{j},t)$ in the 5--7~keV band in the simple lamp-post case. The single source height is varied to be 2, 5 and $10r_{\rm g}$ showing in the black, red and blue lines, respectively. Middle panel: resulting PSD$_{rev}$ (solid lines) overlaid with corresponding PSD$_{0}$ (grey dotted lines). The data are offset for clarity. Bottom panel: the PSD$_{rev}$/PSD$_{0}$ which is the ratio of the full over the excluded-reverberation PSD. The bumps and dips in the ratio PSD are interpreted as reverberation signatures. }
    \label{p1}
\end{figure}

To illustrate this, we present the examples of the simulated light curves, corresponding PSD$_{rev}$, PSD$_0$ and their ratio in Fig.~\ref{p1}. We first set $B_{2}=0$ so the two-blob model reduces to the lamp-post model when only one, lower X-ray source is significant. From now on, if not stated, we fix the inclination $i=30^{\circ}$, the black hole spin $a=0.998$ and produce the light curve in the 5--7~keV band. The black hole mass is fixed at $10^{7}M_{\odot}$. 

Note that the source-response functions are modelled as cut-off power laws (see Fig.~\ref{p-sres}) and the primary variation, $x_{0}(t)$, is generated from a Gaussian distribution, {\tt gsl\_ran\_gaussian()}. Using a top-hat function with different widths for $x_{0}(t)$ is also investigated. The corresponding PSD$_{rev}$ and PSD$_0$ can be noisy and changed according to the choices of $x_{0}(t)$, but their ratio is almost identical for all cases. Moreover, when the model reduces to the lamp-post case (Fig.~\ref{p1}), the first dip, with maximum amplitude, in the ratio PSD appears at lower frequency for higher source height. The maximum amplitude of the first dip also decreases with the source height. These results are well consistent with those in \cite{Papadakis2016} so our approach should be plausible enough for investigating more complex scenarios. 

\begin{figure}
    \centerline{
        \includegraphics[width=0.45\textwidth]{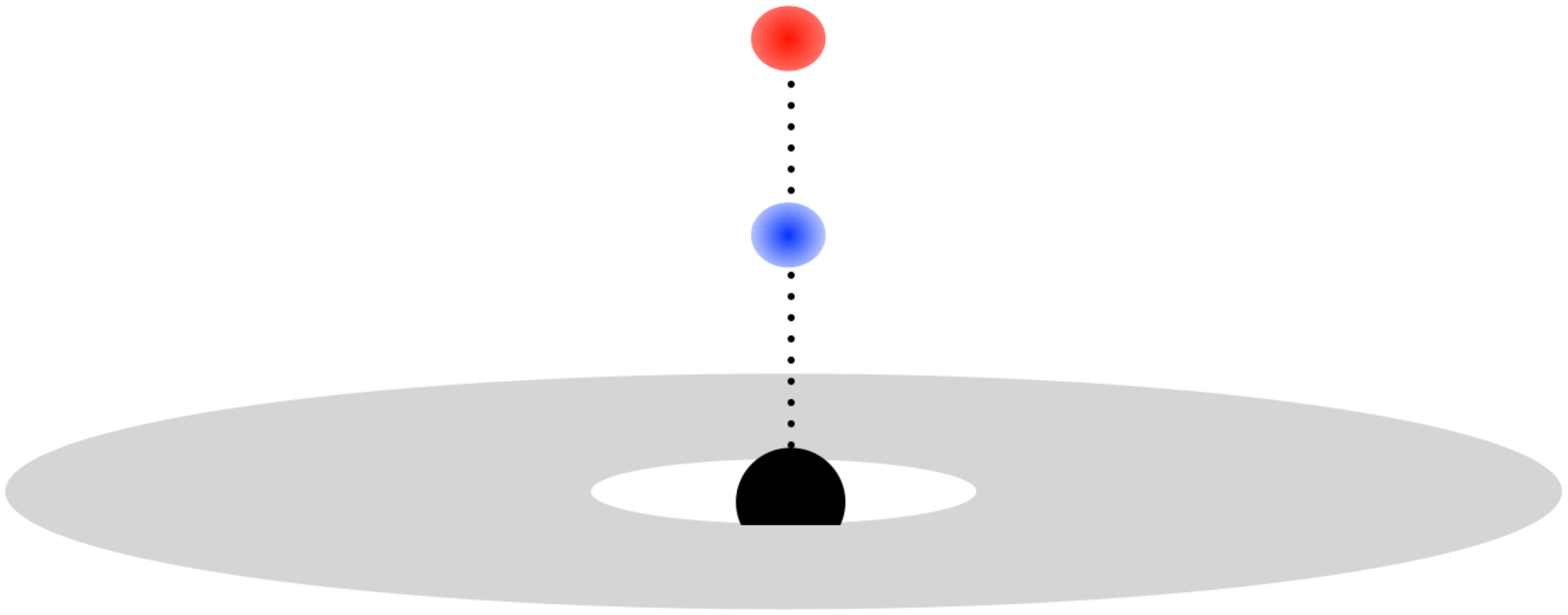}
    }
    \centerline{
        \includegraphics[width=0.45\textwidth]{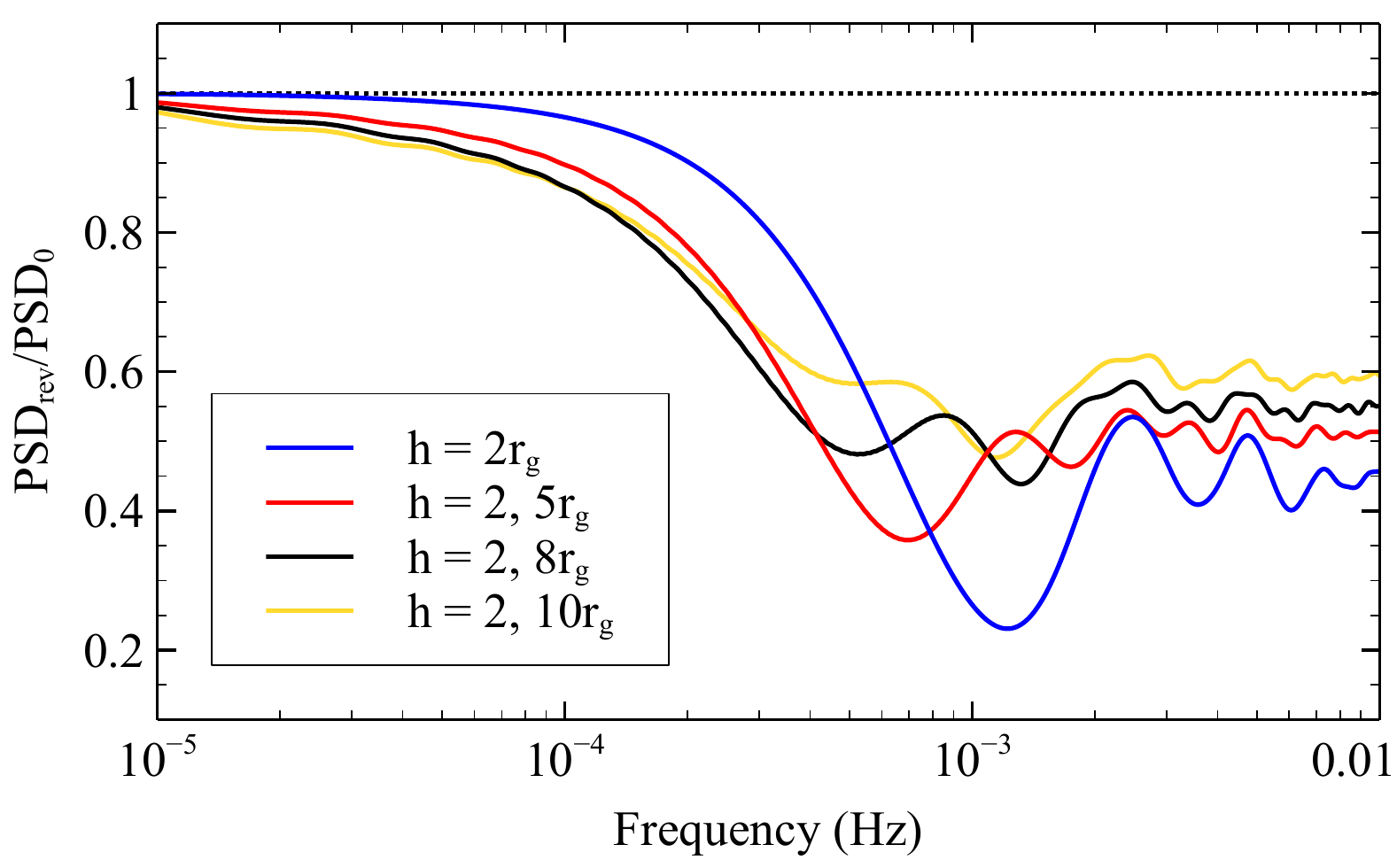}
    }
    \caption{Sketch of the two-blob model (top panel) and corresponding ratio PSD (bottom panel) in the 5--7~keV band varying with the source height. We fix the first, lower blob at $2r_{\rm g}$. The blue line represents the case when the accretion disc is illuminated by only that blob. The PSD when the second blob is located at 5, 7 and $10r_{\rm g}$ are shown in solid red, black and yellow. }
    \label{p2}
\end{figure}

The modelled PSD when the accretion disc is simultaneously illuminated by two X-ray blobs is shown in Fig.~\ref{p2}. We assume $B_{1}=B_{2}=1$ and fix the first source at $2r_{g}$ while placing the second source upper. The continuum emission from the lower and upper source has the photon index $\Gamma_{1}=2$ and $\Gamma_{2}=2.2$. The co-existence of two X-ray blobs leads to a more complex shape of PSD comparing to the lamp-post cases. If the difference of the source height is large enough, the strongest dip in the PSD can be the second dip, rather than the first one (see black and yellow lines in Fig.~\ref{p2}). This is because the upper source produces its first dip at low frequency while the lower source produces its first dip at relatively high frequency with relatively large amplitude. Note that in the lamp-post model the maximum depression in PSD is always seen at the first dip. Therefore, the strongest dip if not the one at the lowest frequency should be evidence supporting reverberation from an extended corona such as a two-blob source that also provides hints to its detailed size and geometry.

\subsection{Two blobs with different brightness}

\begin{figure}
    \centerline{
        \includegraphics[width=0.45\textwidth]{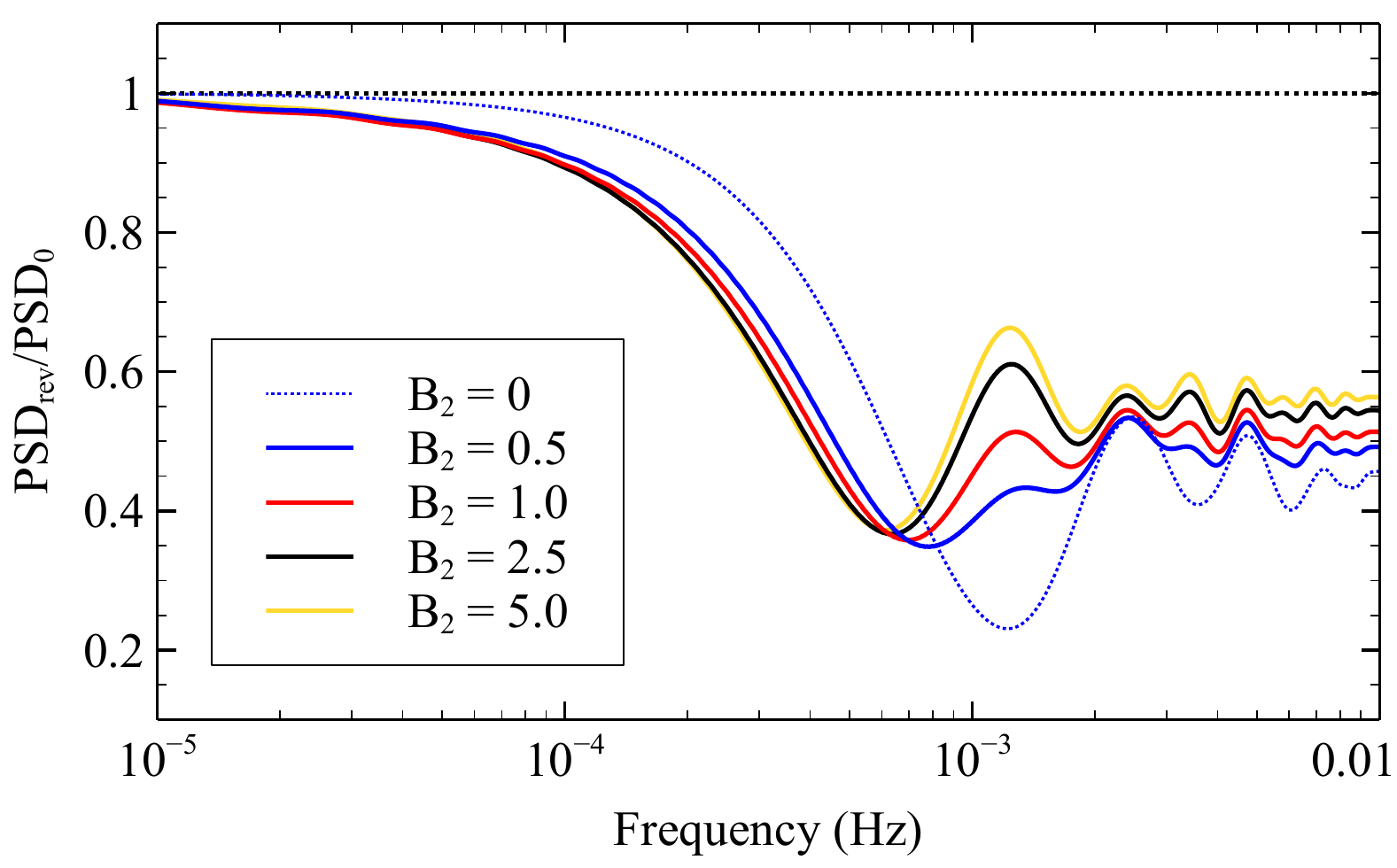}
    }
    \caption{Corresponding ratio PSD (bottom panel) when varying the brightness of the upper source. We fix the blobs at 2 and $5r_{\rm g}$ on the symmetry axis. The blue, red, black and yellow lines represent the cases when the brightness $B_{2}=0.5$, 1.0, 2.5 and 5.0, respectively. We also show the case when $B_{2}=0$ (blue dotted line) that is when the model reduces to the lamp-post case where there is only one x-ray source located at $2r_{\rm g}$.}
    \label{p3}
\end{figure}

From now on we fix the first and the second source at 2 and $5r_{g}$, respectively. The photon index is $\Gamma_{1}=2$ and $\Gamma_{2}=2.2$. Fig.~\ref{p3} shows the ratio PSD when we vary the brightness of the upper X-ray source. Note that the red lines in Fig.~\ref{p2} and Fig.~\ref{p3} are of the same geometry and equal brightness. It can be seen that increasing $B_{2}$ has very small effects on the central frequencies of the dips. Contrarily, larger $B_{2}$ leads to more enhancement of the first hump. Therefore, under the assumption that a physical corona is extended, if the first dip is well determined, the amplitude of the first hump is then capable of providing a major hint towards the locations of two blobs and their relative brightness.  

\subsection{Outflowing and dilution}

\begin{figure}
    \centerline{
        \includegraphics[width=0.45\textwidth]{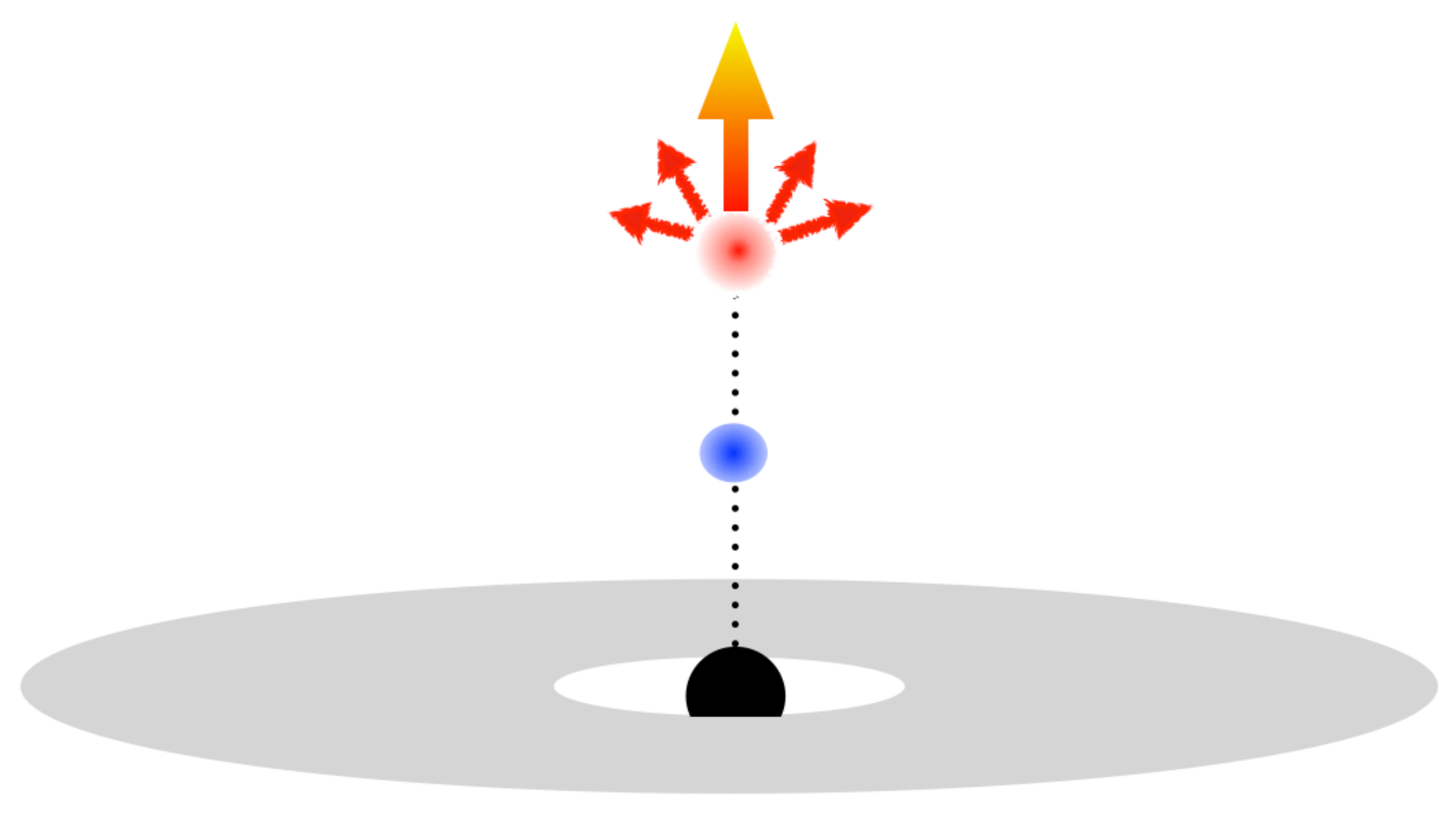}
    }
    \vspace{0.6cm}
    \centerline{
        \includegraphics[width=0.45\textwidth]{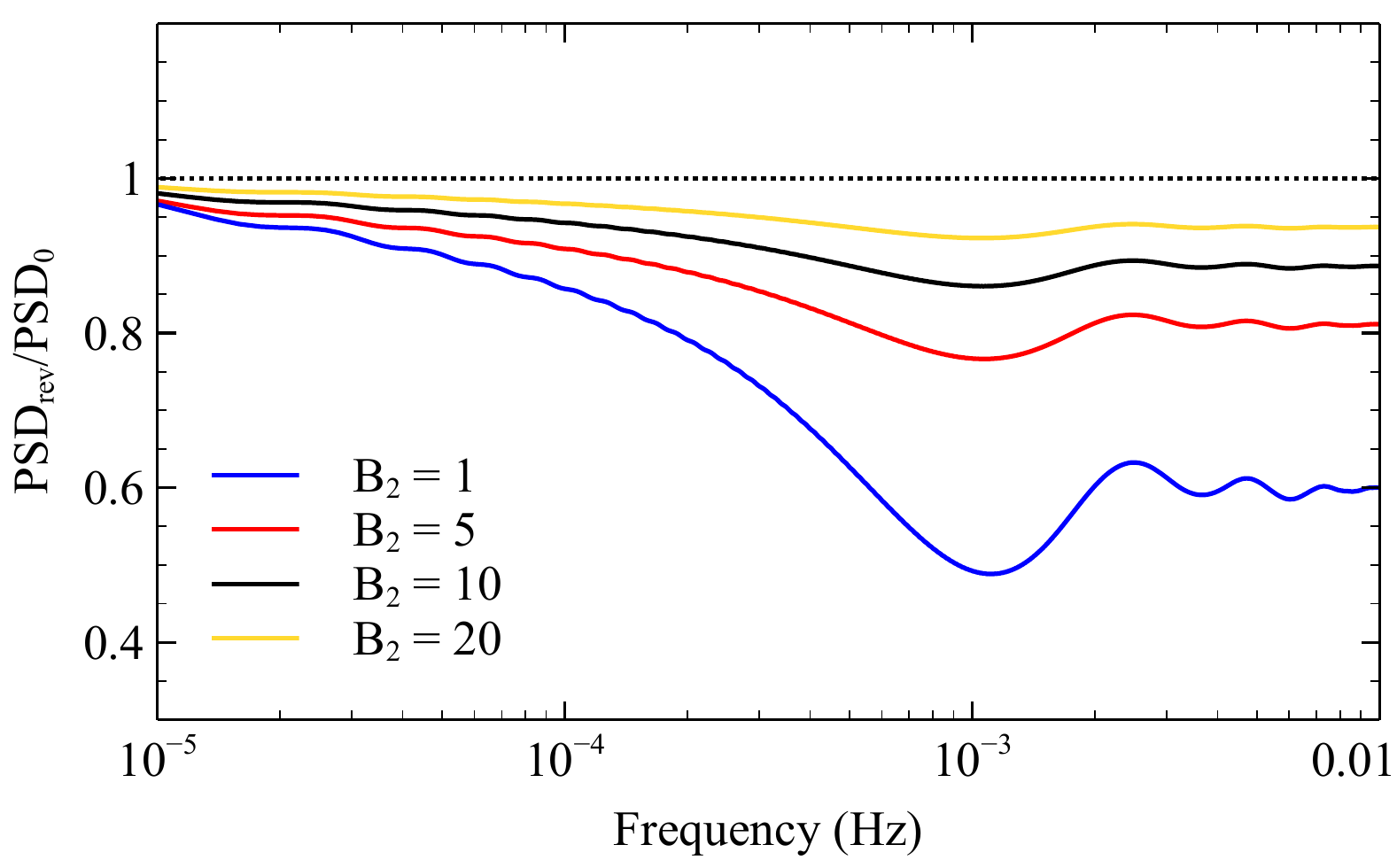}
    }
    \caption{Sketch of the outflowing-blob model (top panel) and corresponding ratio PSD (bottom panel). We fix the lower blob at $2r_{\rm g}$ while the upper blob is outflowing away with different brightness. }
    \label{p5}
\end{figure}

Now we investigate the case when the upper source is outflowing so that its emission is beamed away from the accretion disc. While the photons from the lower X-ray source are traced and the relativistic effects including light bending are taken into account, the upper X-ray source is assumed to move away very fast and produce no reflection. At this stage we simply increase the luminosity of the second source, without allowing this flux to reach the disc (i.e., by setting $r_{2}=0$ in equation~\ref{eq:var2}). This is a crude approximation of the physical conditions in the case of a relativistic jet in AGN. The PSD results are shown in Fig.~\ref{p5}. We can see that the central frequency of each dip is almost identical in all cases and is independent of the parameter $B_{2}$, but the amplitude of the dip is smaller for larger $B_{2}$. This is because when the second source does not contribute to reflection, its continuum plays a role as to dilute the reverberation features. The dip amplitude then associates to the amount of dilution and the reverberation effects are weaker for a brighter relativistic outflows. 

Although the result is not shown here, we find that putting two blobs as the sources of reverberation and place the third source being the outflows that do not contribute the reflection flux into the light curve, the ratio PSD changes in similar trend. This suggests that the strong dilution caused by one high-flux source can be analogically similar as to dilute by the signals from multiple low-flux sources.

\section{Discussion and conclusion} \label{subsubsec:autonumber}

While X-ray reverberation has become a common phenomenon in AGN \citep[e.g.,][]{Kara2016}, the studies of its effects on the PSD of AGN are quite limited, and are likely focused only within the lamp-post scenario \citep[][]{Papadakis2016,Emmanoulopoulos2016}. \cite{Chainakun2017} showed that the bumps and wiggles on reverberation lags seen in recent observations \citep{Caballero-Garcia2018, Caballero-Garcia2019} are naturally produced under the extended corona framework. These studies have drawn attention to the importance of considering the model beyond the standard lamp-post one, and look further to how reverberation manifests in another timing profiles. In principle the accretion disc should have finite thickness and be modulated by mass accretion rate fluctuations. This is, however, the first investigation to explore the effects that reverberation from an extended corona have on the PSD, so for clarity of the results we have chosen to avoid incorporating such complexities that may cause many degeneracies of the model. 

Throughout this work we fix the energy band to be 5--7~keV band. Similar reverberation features should still appear but with smaller amplitude in the band where the reflection component is weaker (e.g., the 2–-4~keV continuum dominated band). The reflection flux can change in response to changes in the ionization state of the disc due to variations of the source luminosity. If the soft-excess component has an inner-disc reflection origin, we expect the reverberation signatures to be strong in the 0.3--1~keV band as well especially in the case of a mildly ionising disc when reflection fraction is large. As suggested by \cite{Papadakis2016}, the soft-excess band may be the best to search for the PSD reverberation features since the signal to-noise ratio is large for current observational data such as from \emph{XMM-Newton}.

Basically, a physical corona could be extended or outflowing \citep{Wilkins2016, King2017, Gallo2019}, and thus oscillatory behaviour of the PSD becomes more complicated comparing to the lamp-post case. The brighter the upper source is, the larger the amplitude of the first hump appears (Fig.~\ref{p3}). If the first dip is well constrained, fine-tuning the model to fit this hump will possibly allow us to probe the relative brightness between the X-ray sources.

Note that the reverberation timescales increase with the source height. The lower source sets the shortest timescales of reverberation and then produces the largest, prominent dip in the PSD profile at high frequency. Contrarily, the upper source sets the longest timescales and hence produces smaller reverberation dip first seen at the lowest frequency. If the corona is extended, the maximum drop in the amplitude of the PSD will not always the first dip seen at the lowest frequency. When the two axial sources are separated far enough, the distinction between the frequencies of which the first dip and the strongest dip appeared become prominent (Fig.~\ref{p2}). 

Although our model is under two-blob hypothesis, these unique oscillator behaviours should be also true to the multiple-source case as they are results from combining multiple response functions, and dilution effects do not change the central frequencies of these dips. We conclude that the highest extent of the corona sets the lowest frequency where the first dip is observed while the lowest extent set the relatively high frequency where the strongest dip is seen. This unique signature could be used to put constraints to the upper and lower limits of the extended corona. 

The oscillations due to reprocessing echoes in the PSD have been searched for in both 0.5--1~keV (soft excess) and 5--7~keV (iron line) bands in a sample of AGN available in the \emph{XMM-Newton} archives \citep{Emmanoulopoulos2016}. Unfortunately, any indications have not been robustly detected yet even the energy bands they used are reflection dominated and the AGN in their sample are among those with the longest \emph{XMM-Newton} observations. The error of the PSD data is still very large due to a limitation of the signal to noise, so a simple bending power-law model could sufficiently provide a good fit to these current data set. It is worth mentioning that the amplitude of the prominent dip is larger for a lower source height in the lamp-post case. The presence of the second, upper source produces distinct physical signatures in the PSD, but it also further dilutes these reverberation features (e.g., Fig.~\ref{p2}). As a result, placing strong constrains to these features expected by an extended corona model should become even more difficult with the present observations.  

Nevertheless, the non-detection of clear oscillatory behaviour may be indicative of an extended corona as well. Our model suggests that a vertical extension of the X-ray source can substantially dilute these characteristics of the PSD, especially if the upper source is outflowing. In some AGN such as Mrk~335 that shows strong evidence of a collimated outflow \citep{Wilkins2015, Gallo2019}, the non-detection with the current data then may be what is expected. Of course, these oscillation features are hard to be robustly confirmed, but the trends reported in \cite{Emmanoulopoulos2016} suggested that the most prominent dip in the observed PSD does not clearly appear at the lowest frequency in any individual sources. This is perhaps a tentative evidence of an extended corona from the PSD profiles of AGN, as predicted by our model (Fig.~\ref{p2}). Since these features can be naturally weak due to dilution effects, statistically significant model-fitting requires higher quality data (e.g., increasing the signal-to-noise ratio). Future observations made by \emph{Athena} will be greatly helpful in reducing the PSD uncertainties, and allow us put a much more accurate constraint to the reverberation signatures in the PSD of AGN.

Furthermore, the light curves produced by two different sources can be anti-correlated in the way that one source is brightening while the flux of the other decreases. It is possible to model anti-correlated signals by setting $x_{2}(E_{j},t)=1/x_{1}(E_{j},t)$ in equations~\ref{eq:var2}--\ref{eq:var3}. However, if any signatures by anti-correlated blobs were detected in the observed PSD, they could still be explained by a combination of differences in the second source's flux, or height. Therefore, such differences between correlated and anti-correlated blobs are not significant. Moreover, the oscillatory structure and behaviour can change over time associating to the mean source flux. If the corona is extending upwards resulting in an increase of source flux, the reverberation signatures will be more diluted and appear at lower frequencies due to the vertical extension of the sources. High quality PSD data, together with time lags, will provide possibility to fit with the model, and to self-consistently constrain the disc-corona geometry as well as its dynamics. Thorough explorations of the relationships between the wiggles seen in the lags and the humps and dips seen in the PSD are planned for the future.

\acknowledgments

PC thanks Suranaree University of Technology for support under grant number SUT1-105-61-12-07, and acknowledges useful discussions with Andrew J. Young and Utane Sawangwit. PC would also like to thank the anonymous referee whose feedback significantly improved the quality of this manuscript.


\end{document}